\documentclass[aps,prb,twocolumn,showpacs,preprintnumbers,amsmath,amssymb,superscriptaddress]{revtex4-1}
\usepackage{subfigure}
\usepackage{graphicx}
\usepackage{dcolumn}
\usepackage{bm}
\usepackage{textcomp}
\usepackage{amsmath}
\usepackage[normalem]{ulem} 
\usepackage{epstopdf}
\usepackage{bm}
\usepackage{multirow}
\usepackage[titletoc,title]{appendix}



\newcommand{\ep}{\varepsilon}

\newcommand{\mH}{{\bf H}}

\newcommand{\bG}{{\bf G}}
\newcommand{\bg}{{\bf g}}
\newcommand{\ba}{{\bf a}}
\newcommand{\mI}{{\bf I}}
\newcommand{\mt}{{\bf t}}
\newcommand{\mT}{{\bf T}}
\newcommand{\mU}{{\bf U}}
\newcommand{\rv}{{\bf r}}

\newcommand{\qv}{{\bf q}}

\newcommand{\kv}{{\bf k}}

\newcommand{\kvec}{{\bf k}}
\newcommand{\Kvec}{{\bf K}}

\graphicspath{{figs/}}

\begin{document}

\title{Interlayer Transport through a Graphene / Rotated-Boron-Nitride / Graphene Heterostructure}

\author{Supeng Ge}
\email[Email:~]{supeng.ge@email.ucr.edu}
\affiliation{Department of Physics and Astronomy, University of
California, Riverside, CA 92521-0204}

\author{K. M. Masum Habib}
\altaffiliation[Current affiliation:]{Intel Corp., Santa Clara CA 95054, USA}
\affiliation{Department of Electrical and Computer Engineering, University of
California, Riverside, CA 92521-0204}

\author{Amrit De}
\affiliation{Department of Electrical and Computer Engineering, University of
California, Riverside, CA 92521-0204}

\author{Yafis Barlas}
\affiliation{Department of Physics and Astronomy, University of
California, Riverside, CA 92521-0204}
\affiliation{Department of Electrical and Computer Engineering, University of
California, Riverside, CA 92521-0204}

\author{Darshana Wickramaratne}
\altaffiliation[Current affiliation:]{Materials Department, University of California, Santa Barbara, CA 93106-5050}
\affiliation{Department of Electrical and Computer Engineering, University of
California, Riverside, CA 92521-0204}

\author{Mahesh  R. Neupane}
\altaffiliation[Current affiliation:]{U.S. Army Research Laboratory, RDRL-WMM-G, Aberdeen Proving Ground, Maryland 21005, USA}
\affiliation{Department of Electrical and Computer Engineering, University of
California, Riverside, CA 92521-0204}

\author{Roger K. Lake}
\email[Email:~]{rlake@ee.ucr.edu}
\affiliation{Department of Electrical and Computer Engineering, University of
California, Riverside, CA 92521-0204}

\begin{abstract}
Interlayer electron transport through a graphene / hexagonal boron-nitride (h-BN) / graphene heterostructure
is strongly affected by the misorientation angle $\theta$ of the h-BN with respect to the graphene layers
with different physical mechanisms governing the transport in different regimes of angle, Fermi level, and bias.
The different mechanisms and their resulting signatures in resistance and current are analyzed using two different models, a
tight-binding, non-equilibrium Green function model and an effective continuum model, and the qualitative 
features resulting from the two different models compare well.
In the large-angle regime ($\theta > 4^\circ$), the change in the 
effective h-BN bandgap seen by an electron at the $K$ point of the graphene causes the resistance
to monotonically increase with angle by several orders of magnitude reaching a maximum at $\theta = 30^\circ$.
It does not affect the peak-to-valley current ratios in devices that exhibit negative differential resistance.
In the small-angle regime ($\theta < 4^\circ$), Umklapp processes open up new conductance
channels that manifest themselves as non-monotonic features in a plot of resistance
versus Fermi level that can serve as experimental signatures of this effect. 
For small angles and high bias, the Umklapp processes give rise to two new current peaks on either
side of the direct tunneling peak.
\end{abstract}


\maketitle 

\section{Introduction}
Graphene (Gr), a two-dimensional (2D) material made of carbon atoms arranged in a honeycomb structure,
has excellent electronic, thermal, and mechanical properties that make it a
promising candidate for nanoelectronic devices\cite{novoselova2012,
RevModPhys.81.109}.
2D hexagonal boron nitride (h-BN) has the same 2D honeycomb structure as graphene.
Its lattice constant is closely matched to that of graphene,
and its large  band  gap and good thermal and chemical stability make it
an excellent insulator, substrate, and encapsulating
material for graphene and other
2D materials.\cite{BN_substrate_Hone_NNano10,xuescanning2011}
There have been a number of experimental and theoretical studies of
the in-plane electronic properties of graphene on
h-BN.\cite{Lanzara_SciRep12,
Massive_Dirac_Ashoori_Sci13,
Levitov_ee_int_graph_BN_PRL13,
Zhao_G_BN_lattice_mismatch_JPCM14,
Novoselov_Comm_incomm_G_BN_NatPhys14,
MacDonald_gaps_NatComm15}
In general, in a h-BN graphene heterolayer system, whether grown by chemical vapor
deposition or assembled by mechanical stacking, the graphene will not be
crystallographically aligned with the h-BN.
The misalignment results in a small change in the
in-plane graphene electron
velocity \cite{Zhao_G_BN_lattice_mismatch_JPCM14}.
%

Interest in the effect of misorientation 
on cross-plane transport began with bilayer graphene,
and the first coherent tunneling calculations showed a 16 order of magnitude change in the
interlayer resistance as a function of the misalignment
angle.\cite{Bistritzer_transport_twisted_PRB10}
Including phonon mediated transport reduced the dependence on angle to a few
orders of magnitude.\cite{Avouris_twisted_PRL12}
Replacing the source and drain misoriented graphene sheets with
source and drain misoriented graphite leads resulted in the same angular dependence and
very similar quantitative values for the coherent current.\cite{Habib_APL13}
This demonstrated sensitivity to interlayer misorientation 
motivates us to examine the effect in Gr/BN/Gr devices.

There is also significant interest in
Gr/BN/Gr heterostructures for electronic device applications \cite{
Britnell_graphene_BN_vFET_Sci12,
Vertical_FET_Geim_Novoselov_Nat12,
Feenstra_GIG_pn_JAP12,
britnellresonant2013,
Ponomarenko_JAP2013,
Zhao_SymFET_TED13,
mishchenkotwist-controlled2014,
Greenaway2015,
PhysRevB.87.075424,
Vogel_Feenstra_G_BN_G__APL14,
fallahazadgate-tunable2015,
zhaonegative2015,
Gaskell2015,
Falko_twist_cntl_APL15,
barrera2015,
barrera2014,
Brey2014,
Vdovin2016,
Guerrero-Becerra_RTD_G_BN_G_PRB16,
Kumar_vertFET_APL12}.
Gr/BN/Gr structures display negative differential
resistance (NDR),\cite{mishchenkotwist-controlled2014,hwan2014,fallahazadgate-tunable2015,Brey2014,Vdovin2016,Guerrero-Becerra_RTD_G_BN_G_PRB16,Falko_twist_cntl_APL15}
and theoretical calculations predict maximum frequencies of
several hundred GHz.\cite{Gaskell2015}
The NDR arises from the line-up of the source and drain graphene Dirac cones combined with
the conservation of in-plane momentum.
In one experiment in which plateaus were observed in the current-voltage
characteristics instead of NDR, the experimental results could be matched theoretically
by ignoring momentum conservation.\cite{Vogel_Feenstra_G_BN_G__APL14}
In the theoretical treatments, the focus has been primarily
on the rotation between top and bottom graphene layers and the resulting misalignment of the
Dirac cones \cite{mishchenkotwist-controlled2014,Falko_twist_cntl_APL15,Guerrero-Becerra_RTD_G_BN_G_PRB16}.
Recently, the effect of misalignment of both the BN and the graphene 
layers including
the effects of phonon scattering have been investigated using
the low-angle effective continuum model \cite{Amorim2016,Brey2014}.
%
%

\begin{figure}
\includegraphics[width=3.0in]{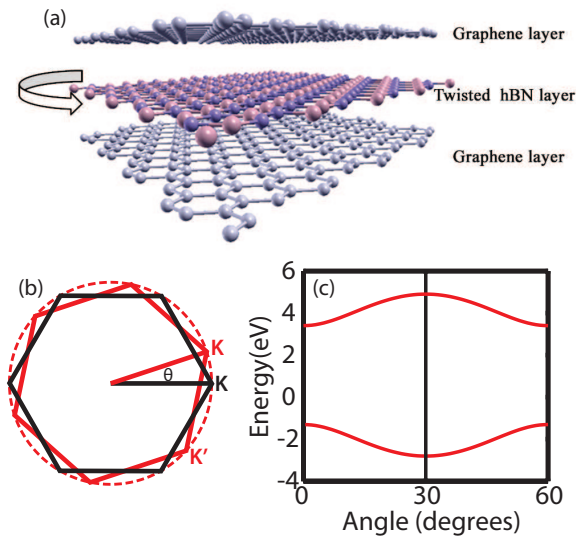}
\caption{
(a) Atomistic geometry of the graphene/boron-nitride/graphene heterostructure.
The top and bottom layers are aligned graphene. 
The middle boron-nitride layer is rotated with respect to the graphene layers.
(b) In $k$ space. The relative rotation between the Brillouin zone of h-BN (red) with respect to that
of graphene (black).
(c) The energy gap of monolayer h-BN at the $K$ point of graphene as a function of rotation angle.
\label{fig:device}
}
\end{figure}

In this work, we focus on the effect of the BN misalignment and consider a system
of two aligned graphene layers serving as the source and the drain
separated by one or more AB stacked layers of h-BN that are misoriented with respect to the
graphene.
An illustration of such a system is shown in Fig. \ref{fig:device}(a).
This system is analyzed using two different models and the results from the two
models are compared.
Commensurate rotation angles in the range $1.89^\circ \leq \theta \leq27.8^\circ$ 
are simulated with
a tight binding model and the
non-equilibrium Green function (NEGF) formalism.
The small angle regime is also analyzed with a continuum model
similar to that used in Ref. [\onlinecite{Amorim2016}].
The qualitative features of the two different models compare well,
and the continuum model elucidates the physics of the
small angle regime 

The misorientation of the BN with respect to the graphene can have several possible effects
that dominate in different regimes of angle and applied bias.
(a) For devices under high bias, it can alter the transverse momentum conservation and thus degrade the NDR.
(b) It can alter the potential barrier seen by the electrons at the K points in the
graphene, and thus alter the interlayer tunneling current and resistance.
(c) As in misoriented graphene on graphene,
it can result in destructive quantum interference that reduces the current.
A signature of this effect is that over a range of angles, the coherent interlayer
resistance scales monotonically with the size of the commensurate
unit cell.\cite{Avouris_twisted_PRL12,Habib_APL13}
(d) For small angle rotations, Umklapp processes can open up new channels of conductance
resulting in new features that depend on Fermi level, angle, and bias.
The presence or absence of these effects and under what conditions they manifest themselves will become 
clear in the analysis.

%

%

The paper is organized as follows. Sec. \ref{sec:TB_NEGF},
describes the tight binding model and the NEGF method used to calculate the
coherent resistance for different commensurate angles and
different h-BN layer thicknesses.
Sec. \ref{sec:eff_cont_model} 
describes the effective continuum model employed to analyze the low angle regime.
Sec. \ref{sec:Results} describes and discusses the results.
Conclusions are given in Sec. \ref{sec:summary}.
The appendix gives details of the tight-binding model and calculations.

\section{Models and Methods} \label{sec:Models}
\subsection{Tight Binding Transport Calculations}\label{sec:TB_NEGF}

The interlayer transport in the Gr/BN/Gr device illustrated in Fig. \ref{fig:device}
is analyzed using a tight binding Hamiltonian
and a non-equilibrium Green function (NEGF) approach for the transport.
%
The device Hamiltonian has the following block tridiagonal form
\begin{equation}
\mH = \begin{pmatrix}
\mH_{T}(\kv) & \mt_{T}(\kv)&0\\
\mt_{T}^\dagger(\kv) & \mH_{BN}(\kv) &\mt_{B}(\kv)\\
0& \mt_{B}^\dagger(\kv) & \mH_{B}(\kv)\\
\end{pmatrix},
\label{eq:H}
\end{equation}
where $\kv$ is the wavevector in the $x-y$ plane,
$\mH_{T(B)}$ is the Hamiltonian of the uncoupled top (bottom) graphene layers,
$\mH_{BN}$ is the Hamiltonian of the h-BN layer(s),
and
$\mt_{T(B)}$ is the block of matrix elements coupling 
$\mH_{T(B)}$ to $\mH_{BN}$.
The elements $t_{ij}$ of the off-diagonal blocks $\mt_{T(B)}$ representing
the interaction between atom $i$ in a graphene layer and atom $j$ in
the adjacent h-BN layer are given by \cite{Avouris_twisted_PRL12}
\begin{equation}
t_{ij} = t_\perp \exp\left(-\frac{r_{ij} - d_{\perp}}{\lambda_z}\right)
\exp\left[\left(\frac{\xi_{ij}}{\lambda_{xy}}\right)^\alpha \right]
\label{eq:tij}
\end{equation}
where $d_{\perp}$ is the interlayer disatnce,
$r_{ij}$ is the distance between two atoms $i$ and $j$,
and $\xi_{ij}=\left[\left(x_i-x_j\right)^2+\left(y_i-y_j\right)^2\right]^{1/2}$
is the projected in-plane distance between the two atoms.
The lattice constant of the entire system is set to that of graphene.
The misoriented commensurate primitive unit cells are created using the approach described 
in Ref. [\onlinecite{Shallcross_band_turbo_graphene_PRB10}].
Parameters for this tight binding model were extracted by fitting the band structures to
density functional theory (DFT) results. 
The on-site energy for $C$, is set to $0$ and the on-site energies of the $B$ and $N$ atoms are 3.40 eV and -1.31 eV, respectively.
For multiple h-BN layers, we adapt the interlayer h-BN interaction 
strength $t'=0.60eV$ from Ref. [\onlinecite{PhysRevB.83.235312}].
All other parameters are shown in Table \ref{tab:PARA1}.

\begin{table}
\centering
\begin{tabular}{ccccccc}
\hline
 & in-plane interaction & \multicolumn{5}{ c }{Interlayer interaction}\\
\hline
 & $t_0\left(eV\right)$&$t_\perp\left(eV\right)$&$d_\perp \left({\rm \AA}\right)$&$\lambda_z\left({\rm \AA}\right)$&$\lambda_{xy}\left({\rm \AA}\right)$&$\alpha$\\
\hline
C-C&2.85&0.39&3.35&0.60&1.70&1.65\\
\hline
B-N&2.52&0.60&&&&\\
\hline
C-B&&0.62&3.22&0.54&0.84&2.04\\
\hline
C-N&&0.38&3.22&0.41&0.97&2.03\\
\hline
\end{tabular}
\caption{Parameters for the tight binding model. $t_0$ is the
intra-layer, off-diagonal matrix element. All other
parameters are described by Eq. (\ref{eq:tij}).
}
\label{tab:PARA1}
\end{table}
Since this is essentially a 2D - 2D tunneling problem,
the coherent interlayer transmission through the Gr/BN/Gr structure is calculated within a NEGF
approach using the `generalized boundary conditions' described in Ref. [\onlinecite{Klimeck_APL95}].
Within the NEGF approach, the graphene layers act as the `contacts' and the h-BN layer
acts as the `device'.
The surface Green's functions of
the top and bottom graphene layers are
\begin{equation}
{\bg}_{T(B)}(E,\kv) = \left[ \left(E + i\frac{\gamma}{2}\right)\mI  - \mH_{T(B)}(\kv) \right]^{-1}
\label{eq:g11}
\end{equation}
where $\mI$ is the identity matrix, and
the energy broadening $\gamma$ = 80 meV
is chosen to match that of Ref. [\onlinecite{bistritzermoiré2011}].
Given the surface Green's functions, the rest of the NEGF calculations follow as usual.
Here the `device' Green's function is
\begin{equation}
\bG^{r}(E,\kv) =
\left[ E\mI - \mH_{BN}(\kv) - {\bm \Sigma}_{T}(E,\kv) - {\bm \Sigma}_{B}(E,\kv) \right]^{-1}
\label{eq:G22}
\end{equation}
where the self energies resulting from coupling to the graphene layers are
${\bm \Sigma}_{T} = \mt_{T}^{\dagger} \bg_{T} \mt_{T}$
and ${\bm \Sigma}_{B}=\mt_{B} \bg_{B} \mt_{B}^{\dagger}$.
The transmission coefficient is
\begin{equation}
T\left(E,\kv\right) = {\rm tr}\left[{\bm \Gamma}_{T} \bG^{r} {\bm \Gamma}_{B} \bG^{r\dagger}\right]
\label{eq:TEK}
\end{equation}
where ${\bm \Gamma}_{T} = \mt_{T}^{\dagger} \ba_{T} \mt_{T}$,
${\bm \Gamma}_{B}=\mt_{B} \ba_{B} \mt_{B}^{\dagger}$,
$\ba_{T(B)} = -i (\bg_{T(B)} - \bg_{T(B)}^\dagger )$ is the spectral function
of the top (bottom) graphene layer,
and ${\rm tr}\left[ \cdots \right]$ indicates a trace of the matrix.

Integrating Eq. (\ref{eq:TEK})
for the transmission over the first commensurate Brillouin zone,
the energy-dependent transmission coefficient per unit area is
\begin{equation}
T(E) = \int_{\rm 1^{st} BZ_c}  \frac{d^2 \kv}{4\pi^2}~ T(E,\kv)
\label{eq:TE}
\end{equation}
This integration is performed numerically on a square grid with
$\Delta k_x = \Delta k_y = 0.005$ ${\rm \AA}^{-1}$
(see Appendix \ref{app:TB_Model_Methods} for further details).
The linear conductance is given by
\begin{equation}
G =  2\frac{\rm e^2}{h}\int dE
T(E)\left(-\frac{\partial f}{\partial E}\right)
\label{eq:R}
\end{equation}
where the factor of 2 accounts for the spin degeneracy, 
and the integration over $\kv$ accounts for the valley degeneracy.
The resistance is the inverse of the conductance, $R=1/G$.

For finite bias calculations, 
an applied bias $V_b = \Delta/e$ is symmetrically applied across the device by setting $\mH_{T}^{i,j}=\delta_{ij}\Delta/2$
and $\mH_{B}^{i,j}=-\delta_{ij}\Delta/2$.
When multiple BN layers are present, the potential drops linearly within the BN region, since BN is an insulator.
The tunneling current flowing through the device is given by:
\begin{equation}
I=\frac{2e}{h}\int dE T(E) \left[ f \left(E - \mu_T \right) - f\left(E - \mu_{B} \right) \right]
\label{eq:I}
\end{equation}
where $\mu_T = \mu_{t} + \Delta/2$ and $\mu_B = \mu_{b} - \Delta/2$ 
are the chemical potentials of the top and bottom graphene, respectively, 
$f(E)$ is the Fermi distribution function, 
and $\Delta V = \mu_{t}-\mu_{b}$ 
is the potential difference between the charge neutral points of the two Gr layers. 
$\Delta V$ accounts for the effect of gating and doping.
We refer to $\Delta V$ as the built-in potential in analogy with a pn junction,
since this is the potential that exists before the bias is applied.

\subsection{Effective Continuum Model} \label{sec:eff_cont_model}

As the rotation angles become smaller the commensurate unit cells become very large. 
As a result, NEGF calculations with the large tight binding Hamiltonians become computationally challenging.
In order to better understand the physics governing the 
interlayer transport at small rotation angles, 
we construct an effective continuum model. 
In the small angle region ($\theta<10^\circ$), 
the coupling matrix between graphene and h-BN layer is of the following 
form \cite{Brey2014,bistritzermoiré2011,mishchenkotwist-controlled2014}
\begin{equation}
\mH_{int} = \frac{1}{3}\sum_{j=1,2,3}e^{-i\qv_i(\theta) \cdot \rv} {\bf T}_j,
\label{eq:Hint}
\end{equation}
where 
\begin{equation}
{\bf T}_j=\begin{pmatrix}
t_{CB} \eta^{(j-1)} & t_{CN}\eta^{-(j-1)} \\
t_{CB}  &  t_{CN} \eta^{(j-1)}  \\
\end{pmatrix}.
\label{eq:Tj}
\end{equation}
In Eq. (\ref{eq:Tj}),
the row indices correspond to the $A$ and $B$ atom of the graphene,
and the column indices correspond to the $B$ and $N$ atoms of the BN.
The lower off-diagonal element corresponds to a C atom directly over a B atom.
All other elements correspond to a C atom in the center of an equilateral
triangle of B atoms or N atoms.
The hopping amplitudes 
$t_{CB}$ and $t_{CN}$ 
between a C atom and a B or N atom are the same as those listed in Table \ref{tab:PARA1}. 
The phase factors 
$\eta=e^{i(2\pi/3)}$
result from the matrix elements of the Bloch sums evaluated at the $K$ points.
The momentum shift
$\qv_i(\theta)$ is the misalignment between the 
$\Kvec$ point of h-BN and graphene. 
Specifically,
\begin{equation}
\begin{aligned}
\begin{split}
&\qv_1(\theta)=k_D(0,\theta),\\
&\qv_2(\theta)=k_D(-\frac{\sqrt{3}}{2}\theta,-\frac{1}{2}\theta),\\
&\qv_3(\theta)=k_D(\frac{\sqrt{3}}{2}\theta,-\frac{1}{2}\theta),
\end{split}\\
\end{aligned}
\label{eq:qvector}
\end{equation}
where $k_D=\frac{4\pi}{3a}$ is the magnitude of the $K$ point of graphene.
When $\theta = 0$, $\qv = 0$, and the sum in Eq. (\ref{eq:Hint}) will cause the 
diagonal and upper off-diagonal
elements of ${\bf H}_{int}$ to vanish leaving a coupling matrix corresponding to
AB stacking with the B atom directly above the C atom.

By eliminating $H_{BN}$ from Hamiltonian (\ref{eq:H}),
we reduce the $3\times3$ Hamiltonian into an effective $2\times2$
Hamiltonian and obtain the effective interaction Hamiltonian between 
the top and bottom graphene layers as \cite{PhysRevB.86.115415}
\begin{equation}
\mU_{TB}(\epsilon) = \mH_{int}(\epsilon-\mH_{BN})^{-1} \mH_{int}^\dagger .
\label{eq:Vint}
\end{equation}
The low-energy electronic structure of h-BN can be described by a gapped 
Dirac-like Hamiltonian that acts on the B and N $p_z$ orbital basis
around a given $K$ point,
\begin{equation}
\mH_{BN}(\kvec) = \begin{pmatrix}
\epsilon_B & \hbar\upsilon_{BN}ke^{i\theta_\kvec} \\
\hbar\upsilon_{BN}ke^{-i\theta_\kvec} & \epsilon_N \\
\end{pmatrix} .
\label{eq:HBN}
\end{equation}
The energies $\epsilon_B$ and $\epsilon_N$ are the on-site energies of the B and N atoms,
while $\upsilon_{BN}$ is the velocity that is determined by the in-plane matrix
elements between the B and N atoms given in Table \ref{tab:PARA1}.
Combining Eqs. (\ref{eq:Hint}), (\ref{eq:Vint}), and (\ref{eq:HBN}), 
the effective interaction Hamiltonian is 
\begin{equation}
\mU_{TB}(\epsilon) = \frac{1}{9}\sum_{i,j=1,2,3}e^{i\bG_{ij}(\theta_T,\theta_B)\rv}\mT_i (\epsilon-\mH_{BN})^{-1}\mT_j
\label{eq:Vint1}
\end{equation}
where $\bG_{ij}(\theta_T,\theta_B)=\qv_i(\theta_T)-\qv_j(\theta_B)$ is the momentum difference shift during transmission.
Since the top and bottom graphene layers are aligned ($\theta_T=\theta_B$),
\begin{equation}
\small{
|G_{ij}|=\left\{
\begin{array}{ll}
0 & \mbox{for $i=j$} \\
\sqrt{3}k_D\theta_{T} & \mbox{for $i\neq$j} \\
\end{array}%
\right.}   \label{Gij}
\end{equation}%
This can be interpreted as the momentum being conserved for transmission between 
aligned Dirac cones of the top and bottom graphene layers.
For transmission between misaligned Dirac cones, the momentum shifts by $|G_{ij}|=\sqrt{3}k_D\theta_{T}$. 

The tunneling matrix element for the transmission between the top and bottom layers is:
\begin{equation}
T_{\alpha,\beta}(\kvec_T,\kvec_B)=\sum_{i,j=1,2,3}t_{i,j}^{\alpha,\beta}(\kvec_T,\kvec_B)\delta(\kvec_T-\kvec_B-\bG_{ij})
\label{eq:Tmatrix}
\end{equation}
where
\begin{equation}
t_{i,j}^{\alpha,\beta}(\kvec_T,\kvec_B)=\frac{1}{9}\phi^\dagger_{\alpha}(\kvec_T)\mT_i (\epsilon-\mH_{BN})^{-1}\mT_j\phi_{\beta}(\kvec_B)
\label{eq:t_ij}
\end{equation}
and the eigenvectors of the graphene layers are
$\phi_{\alpha}(\kvec)=\frac{1}{\sqrt{2}}\left[1, \alpha e^{i\theta_{\kvec}}\right]e^{i\kvec \cdot \rv}$,
where $\alpha=\pm1$ is the band index.
The linear conductance is \cite{Bistritzer_transport_twisted_PRB10}
\begin{flalign}
G = \frac{e^2g_sg_v}{\hbar \cal{A}}
\sum_{\substack{\kvec_{T},\kvec_{B}\\ \alpha,\beta}}
|&T_{\alpha,\beta}(\kvec_T,\kvec_B)|^2 \times
\nonumber \\
&A(\epsilon_\alpha(\kvec_T),\epsilon_F)A(\epsilon_\beta(\kvec_B),\epsilon_F)
&&
\label{eq:G0}
\end{flalign}
or
\begin{flalign}
G =\frac{e^2g_sg_v}{\hbar \cal{A}}
\sum_{\substack{\kvec,\alpha,\beta\\ i,j=1,2,3}}
&|t_{i,j}^{\alpha,\beta}(\kvec,\kvec+\bG_{ij})|^2 \times 
\nonumber \\
&A(\epsilon_\alpha(\kvec),\epsilon_F) A(\epsilon_\beta(\kvec+\bG_{ij}),\epsilon_F)
&&
\label{eq:G}
\end{flalign}
where $g_s=2$ and  $g_v=2$ account for the spin and valley degeneracy, respectively, 
and $\cal{A}$ is the cross sectional area.
$A$ is the spectral function. 
For simplicity we can approximate $A$ by a Lorentzian function near the Fermi 
energy and use a broadening lifetime same as the NEGF calculations\cite{Guerrero-Becerra_RTD_G_BN_G_PRB16}.
%

To better understand the effect of the rotation, we divide the conductance into three parts.
\begin{equation}
G=G_{i=j}+G^{\alpha=\beta}_{i\neq j}+ G^{\alpha\neq\beta}_{i\neq j}
\label{eq:3Gs}
\end{equation}
where the first part
\begin{equation}
G_{i=j}=\frac{e^2g_sg_v}{\hbar}\sum_{\substack{\kvec,\alpha=\beta\\
i=j=1,2,3}}|t_{i,j}^{\alpha,\beta}(\kvec,\kvec)|^2A^2(\epsilon(\kvec),\epsilon_F)
\label{eq:G_direct}
\end{equation}
represents the coherent transport process where the momentum is conserved between top and bottom graphene layers.
The second and third terms correspond to Umklapp processes in which the second term is
an intraband process
\begin{flalign}
G^{\alpha=\beta}_{i\neq j}=\frac{e^2g_sg_v}{\hbar \cal{A}}
& \sum_{\substack{\kvec,\alpha=\beta\\ i \neq j=1,2,3}}
|t_{i,j}^{\alpha,\beta}(\kvec,\kvec+\bG_{ij})|^2 \times \nonumber \\
& A(\epsilon_\alpha(\kvec),\epsilon_F) A(\epsilon_\alpha(\kvec) + \alpha\hbar \upsilon\sqrt{3}k_D\theta,\epsilon_F) ,
&&
\label{eq:G_U_intraband}
\end{flalign}
and the third term is an interband process,
\begin{flalign}
G^{\alpha\neq\beta}_{i\neq j}= & \frac{e^2g_sg_v}{\hbar}
\sum_{\substack{\kvec,\alpha\neq\beta\\ i \neq j=1,2,3}}
|t_{i,j}^{\alpha,\beta}(\kvec,\kvec+\bG_{ij})|^2 \times
\nonumber \\
& 
A(\epsilon_\alpha(\kvec),\epsilon_F ) 
A(\epsilon_\beta (\kvec ) + \beta (\hbar \upsilon\sqrt{3}k_D\theta-2\epsilon_F),\epsilon_F ) .
&&
\label{eq:G_U_interband}
\end{flalign}
%

\section{Results} \label{sec:Results}
\begin{figure}
\includegraphics[width=1\columnwidth]{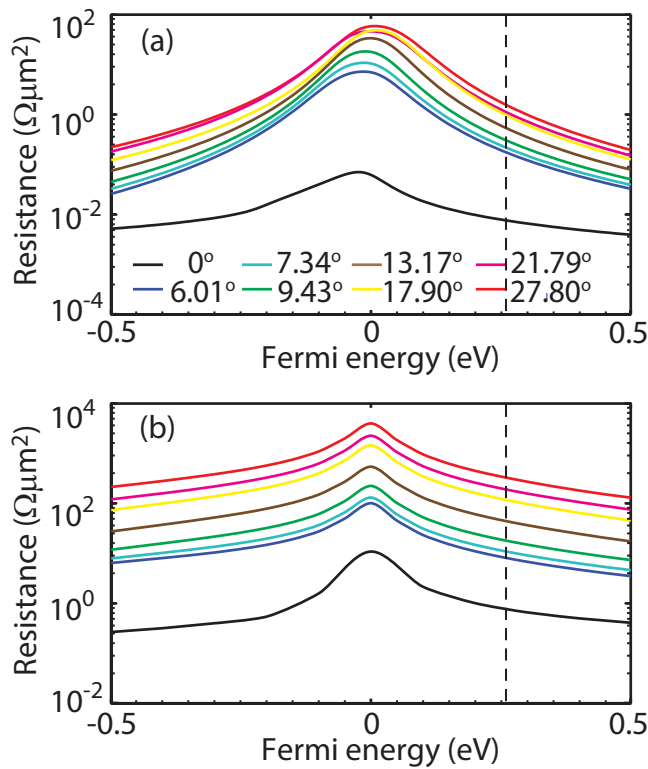}
\caption{
Zero temperature coherent resistance of twisted (a) Gr/1L h-BN/Gr heterostructure and 
(b) Gr/3L h-BN/Gr heterostructure as a function of Fermi Energy for different 
commensurate rotation angles. 
The dashed line shows the Fermi energy of 0.26 eV used to calculate the resistance values in Fig. \ref{fig:TGfitting}.
The resistances are calculated from the tight-binding, NEGF approach.
\label{fig:plotRAll}
}
\end{figure}

Fig. \ref{fig:plotRAll} shows the tight-binding, NEGF calculations of the
zero-temperature, coherent resistance versus
Fermi energy ($E_{F}$) for heterostructures with (a) a single h-BN layer
and (b) 3 h-BN layers.
The Fermi level, $E_F$, varies from -0.5 eV to 0.5 eV around the charge neutrality point
for a range of rotation angles from $0^\circ$ to $27.79^\circ$ as indicated in the legend.
The lowest black curve is the coherent resistance for the ABA unrotated heterostructure.
For all of the angles shown, the resistance monotonically falls as the Fermi level moves away from the 
charge neutrality point where the density of states of the graphene layers are a minimum.
In contrast to rotated bilayer graphene (r-BLG),
for the two lowest angles, $6.01^\circ$ and $7.34^\circ$, 
there is no sudden change in 
resistance with Fermi energy around 0.3-0.4 eV 
(compare with Fig. 2(a-b) of Ref. [\onlinecite{Habib_APL13}]).
The vertical dashed lines in Fig. \ref{fig:plotRAll} correspond to a Fermi level of 0.26 eV.
This is the Fermi level previously used for comparisons of the interlayer conductivity
of misoriented bilayer graphene \cite{Bistritzer_transport_twisted_PRB10,Avouris_twisted_PRL12,Habib_APL13}.
The numerical values of the resistance at $E_F = 0.26$ eV are given in Table \ref{tab:R_GAP}. 
As the h-BN layer becomes misaligned,
the resistances increase by factors of 200 and 430 for 
the monolayer and trilayer BN structures, respectively. 
This trend in the variation of resistance with rotation angle is similar to the 
experimental observations in Ref. [\onlinecite{Britnell_graphene_BN_vFET_Sci12}]. 
There it was shown that the conductance can vary by a factor of
100 for different devices with the same h-BN thickness.
For both the monolayer and trilayer BN structures,
the increase in the resistance is a monotonic function of the BN rotation angle
as the rotation angle increases from $6^\circ$ to $27.79^\circ$.
This trend is also in contrast to that of r-BLG.
In the r-BLG system, at low energies near the charge neutrality point, 
the coherent interlayer resistance is a monotonic function of the 
supercell lattice constant as opposed to the rotation angle 
(compare to Fig. 1(d) of Ref. [\onlinecite{Avouris_twisted_PRL12}]).
\begin{table}
\centering
\resizebox{0.5\textwidth}{!}{
\begin{tabular}{c|c|c|c}
\hline
\multirow{2}{*}{Rotation angle (degrees)} & \multirow{2}{*}{Energy gap (eV)}& \multicolumn{2}{c}{Coherent Resistance ($\Omega  \mu m^2$)}\\
&&Gr/1L h-BN/Gr&Gr/3L h-BN/Gr\\
\hline
\hline
0.00&4.709&0.007601&0.7972\\
\hline
1.25&4.726&0.03710&\\
\hline
1.41&4.730&0.03758&\\
\hline
1.54&4.734&0.03711&\\
\hline
1.61&4.737&0.03521&\\
\hline
1.70&4.740&0.03308&\\
\hline
1.79&4.743&0.03028&\\
\hline
1.89&4.748&0.02844&2.752\\
\hline
2.00&4.752&0.02954&\\
\hline
2.13&4.758&0.03481&\\
\hline
2.45&4.774&0.05355&\\
\hline
2.88&4.798&0.07565&4.474\\
\hline
3.15&4.815&0.08741&\\
\hline
3.48&4.838&0.09981&\\
\hline
3.89&4.869&0.1132&5.510\\
\hline
4.41&4.913&0.1288&6.094\\
\hline
5.08&4.976&0.1481&6.977\\
\hline
6.01&5.075&0.1753&8.495\\
\hline
7.34&5.237&0.2182&11.43\\
\hline
9.43&5.529&0.3048&18.87\\
\hline
13.17&6.106&0.5371&46.48\\
\hline
17.90&6.813&0.9770&123.6\\
\hline
21.79&7.280&1.120&199.7\\
\hline
27.80&7.686&1.563&344.3\\
\hline
\end{tabular}
}
\caption{Effective BN energy gap and the coherent resistances at $E_{F}$=0.26 eV for different 
commensurate rotation angles
and two different BN thicknesses of 1ML and 3ML. The resistances are calculated from the tight-binding, NEGF
approach.
}
\label{tab:R_GAP}
\end{table}

To investigate process (b) in which
rotation of the BN alters the tunnel barrier,
we calculate the energy gap of ML and trilayer h-BN at the BN $k$-point 
corresponding to graphene's $K$-point as a function of rotation angle
as illustrated in Fig. \ref{fig:device}(b). 
The resulting effective bandgap for ML BN is plotted versus
rotation angle in Fig. \ref{fig:device}(c).
Since the direct bandgap (4.7 eV) of h-BN occurs at its $K$-point,
the minimum BN bandgap `seen' by an electron at the $K$-point in the graphene layer
occurs for BN rotation angles of $0^\circ$ and $60^\circ$
when graphene's $K$ point is aligned with BN's $K$ or $K'$ points.
The effective BN bandgap seen by an electron at the $K$-point in the graphene layer
monotonically increases as the BN is rotated from $\theta = 0^\circ$, 
and it reaches a maximum at $\theta = 30^\circ$.
In the Brillouin zone of the BN, this corresponds to the bandgap near the $M$ point.
This monotonic increase in the tunnel barrier with angle 
follows the same monotonic trend as the increase in resistance with angle.

\begin{figure}
\includegraphics[width=1\columnwidth]{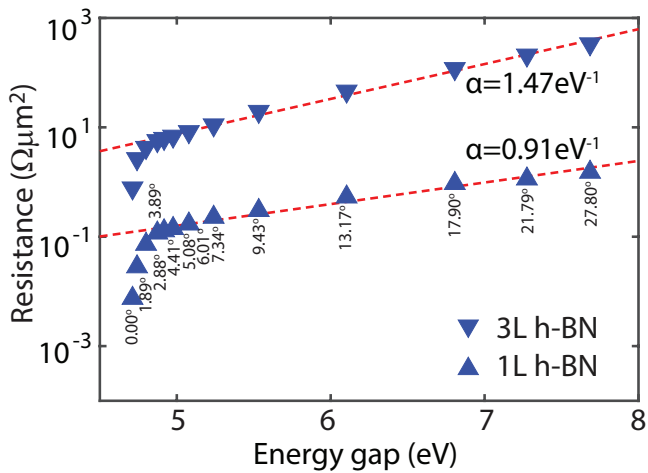}
\caption{
Zero temperature coherent resistance of graphene/1L h-BN/graphene (upward-pointing triangles) 
and graphene/3L h-BN/graphene (downward-pointing triangles) as a function of 
the effective energy gap of monolayer h-BN at the K-point of the graphene. 
The angles are given next to each data point.
The red lines show exponential fits to the data, $R = R_0 e^{\alpha E_G}$. 
The values of $\alpha$ are shown next to the fitted line.
$E_{F}$=0.26 eV. 
\label{fig:TGfitting}
}
\end{figure}

To analyze the relation between the effective energy gap and resistance, 
we show in Fig. \ref{fig:TGfitting} a semi-log plot of the resistance as a function of the effective BN band gap 
(for different rotation angles) at $E_{F}=$0.26 eV.
For angles greater than $4^\circ$, the tunnel current scales exponentially
with the effective bandgap as one would expect for tunneling through a potential barrier.
Therefore, for $\theta > 4^\circ$, we find that the dominant process
affecting the tunnel current is the change in the effective BN bandgap `seen' by
the electrons at the $K$ point in graphene.

However, for small angles $\theta < 4^\circ$, there is clearly a very different trend and
a different dependence of the resistance on the BN rotation angle.
The different dependencies arise from different parallel conductance channels that dominate
at different angle regimes.
To analyze the low-angle region of the curve, we turn to the effective continuum model.

\begin{figure}
\includegraphics[width=1\columnwidth]{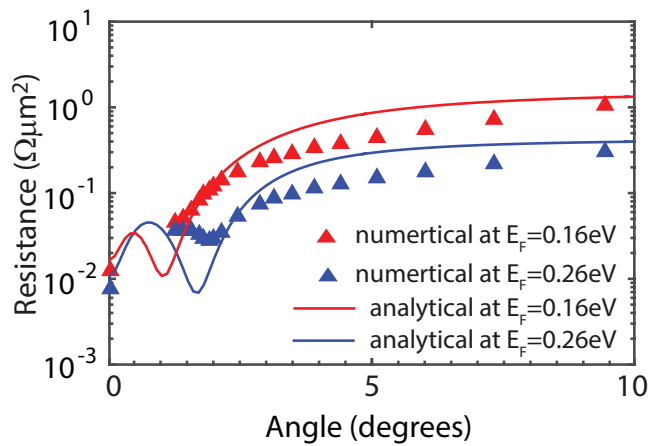}
\caption{
Zero temperature coherent resistance of Gr/1L h-BN/Gr as a function of rotation angle for 
Fermi energies equal to 0.26eV (blue) and 0.16eV (red). 
The solid lines show the result calculated from the continuum model, 
and the triangles show the results from the tight-binding, NEGF calculation.
The smallest commensurate rotation angle calculated numerically is $1.25^\circ$.
\label{fig:R_NEGF_Continuum}
}
\end{figure}

A more detailed picture of the low-angle regime is given in Fig. \ref{fig:R_NEGF_Continuum}
which shows the resistance versus BN rotation angle calculated
with both the continuum model and the NEGF tight-binding model for two values of $E_F$.
The solid lines are from the continuum model, and the triangles are from the NEGF, tight-binding
model.
More low-angles are included in the NEGF calculations, and
the smallest rotated angle calculated from the NEGF, tight-binding model is $1.25^\circ$.  
Both models show a non-monotonic dependence of resistance on angle at very low angles $\theta < 2.5^\circ$. 
While the magnitudes differ between the two models, the overall trends match well.

The continuum model tells us that there are three parallel conductance channels
corresponding to the direct and two Umklapp processes in Eqs. (\ref{eq:3Gs}) - (\ref{eq:G_U_interband}).
The individual channels dominate in different angle regimes. 
%
The angle at which each channel dominates is primarily determined by the overlap of the spectral
functions in Eqs. (\ref{eq:G_direct}) - (\ref{eq:G_U_interband}).
For the direct term, $G_{i=j}$ of Eq. (\ref{eq:G_direct}), 
the spectral functions always overlap
since the top and bottom graphene layers are aligned.
For the two Umklapp terms, the overlaps of the spectral functions
are functions of the angles, and the overlaps
become negligible for $\hbar \upsilon\sqrt{3}k_D\theta>> \hbar/\tau,\; \epsilon_F$.
Therefore, for larger angles, $\theta > 4^\circ$, the direct channel dominates,
and the dependence on the angle is through the matrix element
which, through ${\bf H}_{BN}(\kv)$ and the effective interaction, 
includes the effect of the increase in the apparent BN bandgap with angle as described above
and shown in Fig. \ref{fig:device}(c).

The maximum overlap of the spectral functions in the `interband' term of Eq. (\ref{eq:G_U_interband})
occurs when $\hbar \upsilon\sqrt{3}k_D\theta = 2\epsilon_F$.
This term is maximum at rotation angle $\theta_m = 2\epsilon_F / \hbar \upsilon\sqrt{3}k_D$,
and it decreases for angles greater than or less than $\theta_m$.
This interband term is responsible for the dip in resistance for $\theta$ between one to two degrees
in Fig. \ref{fig:R_NEGF_Continuum}.
It also explains the shift in angle with Fermi level.
As the Fermi level is increased, the local minimum moves to larger rotation angles since the
angle of maximum overlap $\theta_m$ is linearly proportional to $\epsilon_F$.

\begin{figure}
\includegraphics[width=1\columnwidth]{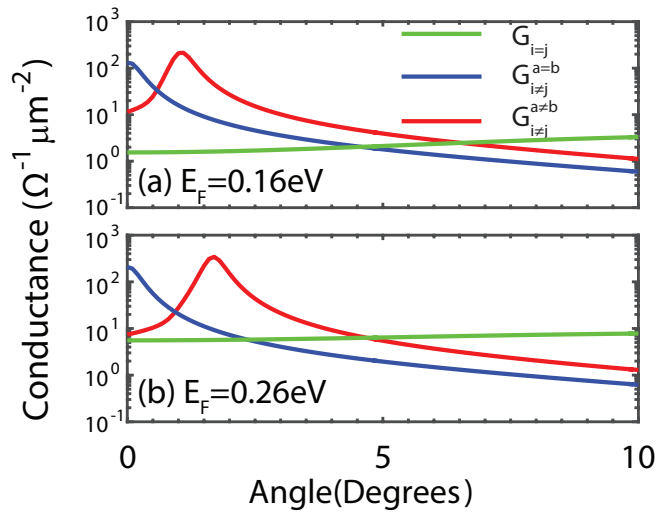}
\caption{Conductance components as a function of rotation angle for (a) $E_F=0.16eV$, (b) $E_F=0.26eV$.
\label{fig:conduct_component}
}
\end{figure}

The maximum overlap of the spectral functions in the `intraband' term of 
Eq. (\ref{eq:G_U_intraband}) occurs at $\theta = 0$.
As $\theta$ increases, 
this channel monotonically decreases with the decrease governed
by the decreasing overlap of the spectral functions. 
Since this channel has a maximum as $\theta$ goes to zero, it governs the initial increase in resistance
for the smallest angles.
The three individual contributions to the continuum model, direct, interband, and intraband, 
are shown in Fig. \ref{fig:conduct_component} for the two different Fermi levels, 0.26 eV and 0.16 eV.

\begin{figure}
\includegraphics[width=1\columnwidth]{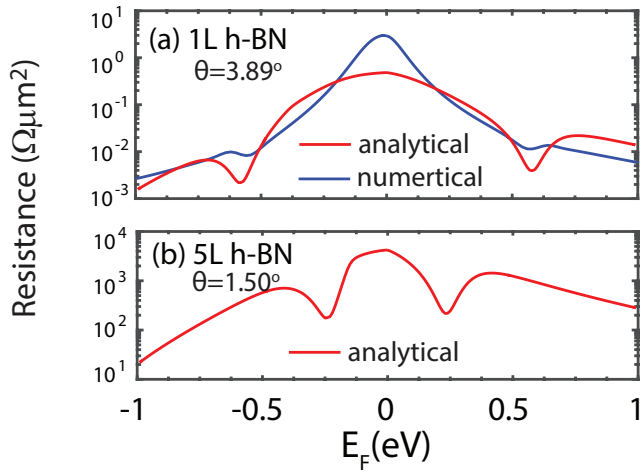}
\caption{Resistance versus Fermi level of the Gr/1L-hBN/Gr structure 
(a) calculated by both the continuum model and the NEFG method with $\theta = 3.89 ^\circ $,
and
(b) calculated by the continuum model only with $\theta = 1.50 ^\circ $.
\label{fig:R_VS_EF}
}
\end{figure}

While analyzing the resistance as a function of rotation angle is useful for clarifying the physics, 
verifying the trends shown in Fig. \ref{fig:R_NEGF_Continuum} would be very difficult experimentally.
Experimentally, it is far easier to fix the angle and sweep the Fermi level of the top and bottom
graphene layers. 
The resulting resistances calculated both from the NEGF, tight-binding and the continuum models 
for a 1-ML BN rotation angle of $3.89^\circ$ are shown in Fig. \ref{fig:R_VS_EF}(a).

Both models show non-monotonic behavior of the resistance as the Fermi level is swept between
0.5 and 0.6 eV.
To observe this feature at lower Fermi levels, a smaller angle is required, and to observe the feature
experimentally a larger resistance is required. 
The larger resistance is achieved by increasing the number of BN layers from 1 to 5. 
The resistance versus Fermi level calculated from the continuum model
for a 5-ML BN layer rotated by $1.50^\circ$ is shown in Fig. \ref{fig:R_VS_EF}(b). 
The non-monotonic feature moves to lower energies and now occurs as the Fermi level is swept between 0.2 and 0.3 eV. 
The overall magnitude of the resistance is between 100 and 1000 $\Omega \mu m^2$ which should be large enough
to be ovservable, and it can be increased by increasing the number of BN layers.

\begin{figure}
\includegraphics[width=3in]{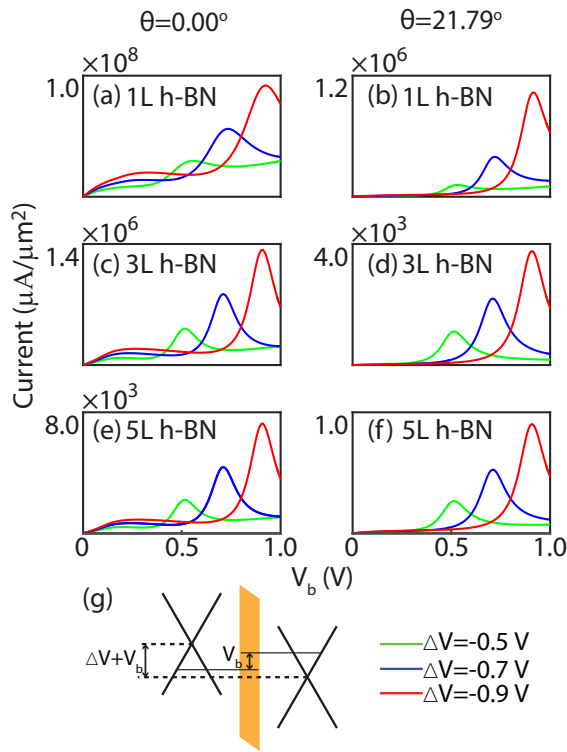}
\caption{
Current as a function of bias voltage for different potential differences $\Delta V$ between the two graphene layers. 
(a) Graphene/1L h-BN/graphene with no rotation; 
(b) graphene/1L h-BN/graphene with a 21.79$^0$ rotation angle; 
(c) graphene/3L h-BN/graphene with no rotation; 
(d) graphene/3layer h-BN/graphene with a 21.79$^0$ rotation angle; 
(e) graphene/5L h-BN/graphene with no rotation; 
(f) graphene/5layer h-BN/graphene with a 21.79$^0$ rotation angle.
\label{fig:IV}
}
\end{figure}

\begin{figure}
\includegraphics[width=3in]{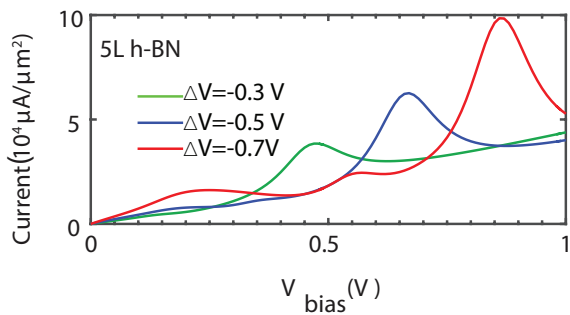}
\caption{
Current as a function of bias voltage for different potential differences $\Delta V$ 
between the two graphene layers for the 5L h-BN structure with a h-BN rotation angle of $\theta=0.5^\circ$
\label{fig:IV_continnum}
}
\end{figure}

So far, we have focused on the 0-bias resistivity to elucidate the physics.
However, interest in this system is driven by potential applications,
and one application of current investigation is a high-frequency oscillator
that exploits the negative differential resistance observed under high-bias.
To understand how the misorientation of the BN layer affects the current-voltage (I-V)
characteristic of this structure, we show in Fig. \ref{fig:IV} the NEGF, tight-binding calculations
using Eq. (\ref{eq:I}) 
of the I-V characteristics for the unrotated structure and the structure with
the BN layer rotated by $21.78^\circ$ for BN layer thicknesses of 1 ML, 3 ML, and 5 ML.
The three I-V characteristics in each plot are for three different built-in potentials
$\Delta V$ between the two graphene layers.
The panels on the left are for the unrotated structure while the panels on the 
right are for the $21.79^\circ$ structure.
In Fig. \ref{fig:IV}(a) and (b), it is shown that the rotation of monolayer h-BN 
decreases the current by nearly 2 orders of magnitude. 
This relative decrease in the tunneling current becomes progressively greater as the 
number of h-BN layers is increased, as shown in the other subplots. 
For the case of 5 h-BN layers, the tunneling current is nearly 4 orders of magnitude smaller. 
As expected, this decrease in the tunneling current and its scaling is consistent with 
the resistance increasing with the rotation angles as shown in Fig. \ref{fig:plotRAll}.
While the current decreases with rotation angle, the peak-to-valley current ratio is unaffected.
For high-frequency applications, both high current density 
and high peak-to-valley ratios are desirable, and rotation of the BN
layer provides one more tool for engineering optimal electronic properties
for applications.
For small rotation angles, it is interesting to consider whether new
qualitative features appear in the nonlinear I-V characteristic.
To answer that question,
we applied the effective continuum model to calculate I-V curves of a structure with $\theta = 0.5^\circ$.
The results in Fig. \ref{fig:IV_continnum}, 
for 3 different values of built-in voltage $\Delta V$, are qualitatively
different from the I-V curves for large angle rotation,
since several regions of NDR appear depending on the initial
built-in potential. 
The first and third peaks arise from the interband component which is maximum 
at $V_{bias}= \pm \hbar \upsilon\sqrt{3}k_D\theta - \Delta V$.
The middle peak that occurs at $V_{bias}=-\Delta V$ is caused by the direct tunneling term.

\section{Conclusions} \label{sec:summary}

Electron transport through a Gr / h-BN / Gr structure is examined within a tight-binding
model with commensurate rotation angles and within an effective continuum model.
The two graphene layers are aligned, and the h-BN layer is rotated by an angle $\theta$
with respect to the graphene layers.
For angles greater than $4^\circ$, the resistance is dominated by
the change in the effective h-BN bandgap
seen by an electron at the $K$ point of the graphene.
In this large-angle regime, the effect of rotating the BN  
is to increase the barrier height of the BN tunnel barrier at the $K$ point of the graphene.
For $\theta \gtrsim 4^\circ$, the resistance monotically increases with the rotation angle, and it
reaches a maximum at $\theta = 30^\circ$.
As $\theta$ is increased from $0^\circ$ to $30^\circ$, the coherent interlayer resistance increases
by factors of 200 and 430 for monolayer and trilayer BN layers, respectively.
For devices that exhibit NDR under high bias, rotation of the h-BN primarily serves
to reduce the overall magnitude of the current.
It does not degrade the peak to valley current ratios.
In this large-angle regime, since the dominant physics is that of single-barrier
direct tunneling, phonon-scattering should have negligible effect on the low-bias, angle-dependent trends
and magnitudes of the interlayer resistances.
Since NDR results from momentum conservation, phonon-scattering will reduce the peak-to-valley
ratios, but this effect also exists in the unrotated structure. 
While we do not expect a significant 
dependence of the phonon scattering on the rotation angle of the h-BN in the large-angle regime,
this is an open question for further study.

The small-angle regime ($\theta \lesssim 4^\circ$) reveals qualitatively new features both in the low-bias
interlayer resistances and in the high-bias I-V characteristics.
The new features arise due to the opening of new conductance channels corresponding to
Umklapp processes.
With the two graphene layers aligned, Umklapp processes give rise to two new
conduction channels corresponding to an intraband term and an interband term.
The angular and energy dependence of these terms is primarily determined
by the overlap of the top and bottom graphene spectral functions that are shifted in momentum space
with respect to each other by an Umklapp lattice vector.
For a fixed rotation angle $\theta$ of the h-BN layer, both the intraband 
and interband terms peak at a Fermi level
$\ep_F^m \equiv \hbar v k_D \theta \sqrt{3} / 2$.
At this Fermi level, the two spectral functions in the interband term perfectly
overlap, so that the interband term dominates.
This strong peak in the interband term results in a distinct, non-monotonic feature
in a plot of the interlayer resistance versus Fermi energy that occurs
as the Fermi level is swept through $\pm \ep_F^m$.
The qualitative trends of this non-monotonic feature are 
reproduced in the tight-binding calculations for structures with small
commensurate rotation angles, although the overall magnitude of the 
feature is less. 
The interband term also gives rise to two extra peaks in the nonlinear $I-V$ characteristic
on either side of the peak resulting from the direct tunneling term.
Amorim et al. \cite{Amorim2016} found that 
phonon scattering and incoherent scattering in this low-angle regime reduces the magnitude
of the features resulting from Umklapp processes, but it does not remove them, so that the
new features in the low-angle regime should be experimentally observable.
\\
\\

\noindent
{\em Acknowledgement}: 
This work is supported in part by FAME, one of six centers of STARnet, 
a Semiconductor Research Corporation program sponsored by MARCO and DARPA
and the NSF EFRI-143395.
This work used the Extreme Science and Engineering Discovery Environment (XSEDE), 
which is supported by National Science Foundation grant number ACI-1053575.

\begin{appendices}
\section{Tight-binding model and method details} \label{app:TB_Model_Methods}
\begin{figure}
\includegraphics[width=1\columnwidth]{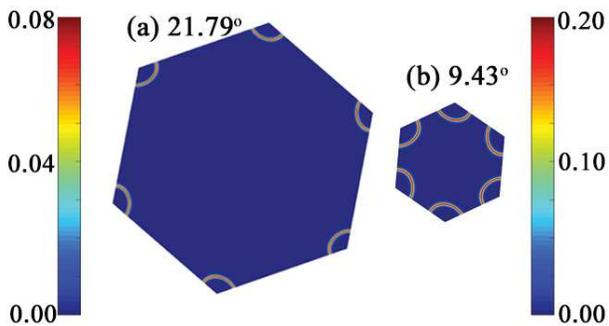}
\caption{Transmission coefficient $T(E,\kv)$ in the first Brrillouin at energy of 0.5 eV for Graphene/1L h-BN/Graphene heterostructure with rotation angel: (a) $21.79^\circ$ (b)$9.43^\circ$
\label{fig:T_BZ}
}
\end{figure}

The transmission coefficient over $\kv$ in the first Brillouin zone, 
$T(E) = \int_{\rm 1^{st} BZ}  \frac{d^2 \kv}{4\pi^2}~ T(E,\kv)$ 
was numerically integrated on a square grid with $\Delta k_x = \Delta k_y = 0.005$ ${\rm \AA}^{-1}$.
Fig. \ref{fig:T_BZ} shows the momentum resolved transmission $T(E,\kv)$ in the 
first Brillouin zone corresponding to the two commensurate rotation angles of $21.79^\circ$
and $9.43^\circ$ at $E = 0.5$ eV.
The transmission is centered at the K and K' and peaks on the isoenergy surface.

\begin{figure}
\includegraphics[width=1\columnwidth]{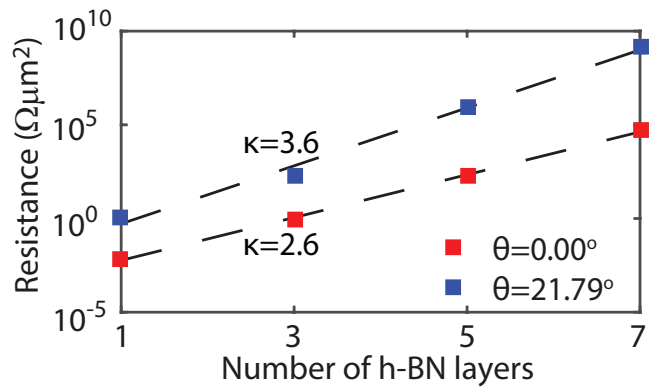}
\caption{Resistance versus number of h-BN layers for rotation angles of $0.00^\circ$ and $21.79^\circ$ 
at a Fermi energy of $E_F=0.26eV$. 
The dash lines show the exponential fits $R = R_0 e^{\kappa \cdot n}$ where $n$ is the number
of BN layers.
The decay constants $\kappa$ are shown next to the fits for the two structures.
\label{fig:R_numb}
}
\end{figure}

To extract a tunneling decay constant of the BN predicted by the interlayer tight-binding parameters,
we calculate the resistance of 1, 3, 5, and 7 layers of h-BN for two angles of $\theta=0$ and $\theta=21.79^\circ$
at $E_F = 0.26$ eV.
Fig. \ref{fig:R_numb} shows the exponential increase in resistance 
with increasing number of h-BN layers for both structures.
Fitting the results to an exponential function, $R = R_0 e^{\kappa \cdot n}$, 
where $n$ is the number of h-BN layers gives values for $\kappa$ of 2.6 and 3.6
for the unrotated and rotated structures, respectively.
These values are similar to an experimentally extracted value of $\kappa=4.0$ \cite{britnellnl3002205}.

\end{appendices}

\newpage

%

\end{document}